\documentclass{apj}

\newcommand{\pol}{polo\"{\i}dal }
\newcommand{\tor}{toro\"{\i}dal }
\newcommand{\be}{\begin{equation}}
\newcommand{\ee}{\end{equation}}
\newcommand{\p}{\partial}
\newcommand{\nab}{{\bf \nabla}}

\newcommand{\JJ}{\mbox{\bf J}}

\newcommand{\ez}{\hat{{\bf e}}_Z}
\newcommand{\er}{\hat{{\bf e}}_R}
\newcommand{\ep}{\hat{{\bf e}}_P}
\newcommand{\BB}{\mbox{\bf B}}
\newcommand{\FF}{\mbox{\bf F}}
\newcommand{\EE}{\mbox{\bf E}}
\newcommand{\vv}{\mbox{\bf v}}

\slugcomment{The Astrophysical Journal, in press}

\shorttitle{Radiatively inefficient MAES} 
\shortauthors{CASSE \& KEPPENS}

\begin{document}

\title{Radiatively inefficient MHD accretion-ejection structures}

\author{Fabien Casse and Rony Keppens}
\affil{FOM-Institute for Plasma Physics Rijnhuizen, Association Euratom/FOM, \\
P.O. Box 1207, 3430 BE Nieuwegein, The Netherlands \\
fcasse@rijnh.nl, keppens@rijnh.nl\\
Submitted, July 18th, 2003; Accepted October 2nd, 2003}

\begin{abstract}
We present magnetohydrodynamic simulations of a resistive accretion
disk continuously launching transmagnetosonic, collimated jets. 
We time-evolve the full set of magnetohydrodynamic equations, but neglect
radiative losses in the energetics (radiatively inefficient). 
Our calculations demonstrate that a jet is self-consistently produced by the
interaction of an accretion disk with an open, initially bent 
large-scale magnetic field. A constant fraction of heated disk
material is launched in the inner equipartition disk regions, leading to
the formation of a hot corona and a bright collimated, super-fastmagnetosonic
jet. We illustrate the complete dynamics of the ``hot''  near steady-state
outflow (where thermal pressure $\simeq$ magnetic pressure) 
by showing force balance, energy budget and current circuits. The
evolution to this near stationary state is analyzed in terms of 
the temporal variation of energy fluxes controlling 
the energetics of the accretion disk. We find
that unlike advection-dominated accretion flow, the energy released by
accretion is mainly sent into the jet rather than transformed into disk
enthalpy. These magnetized, radiatively inefficient 
accretion-ejection structures can account for
under-luminous thin disks supporting bright fast collimated jets as seen in
many systems displaying jets (for instance M87). 
\end{abstract}

\section{Accretion disks and jets}

\subsection{Accretion-Ejection models}
Astrophysical jets are quite common phenomena across our visible 
universe. They are typically observed in association with 
accreting objects such as low-mass young stellar
objects (YSO), X-ray binaries (XRB) or active galactic nuclei (AGN,
e.g. \citet{Livi97} and references therein). In all these systems, the mass
outflows exhibit very good collimation at large distances from the central
object as well as high velocities along the jet axis. Although operative at
widely 
disparate lenghtscales, these accretion-ejection systems have other features
in common, in particular, observational links have been established
in all cases between accretion disk
luminosity and jet emission: for YSO see e.g. \citet{Hart95}, for XRB see
\citet{Mira98} and for AGN \citet{Serj98}.

The most promising unifying model relies on a scenario where
an accretion disk interacts with a large-scale magnetic field in order to
give birth to bipolar collimated jets 
(for the specific case of early protostars see
also \citet{Lery99}). Since the seminal work by~\citet{Blan82}, it is known 
that the action of an open magnetic field configuration threading a disk
can brake rotating matter in order
to transfer angular momentum into the jet and provide energy for
acceleration of jet matter. This magnetohydrodynamic (MHD) model 
describes the interaction of the accretion flow with a magnetic field whose 
origin can be due to advection of interstellar magnetic field \citep{Mous76}
and/or dynamo produced \citep{Reko03}. The collimation of the flow is
self-consistently achieved by the electric current produced by
the flow itself \citep{Heyv89}. This current provokes a radial
pinching of the plasma that can balance both magnetic and thermal pressure
gradients \citep{Saut02}. The analytical work by~\citet{Blan82} did make
the simplifying assumption of a cold plasma.

Numerous studies have dealt with these Magnetized Accretion-Ejection Structures
(MAES) in the last two decades. Sophisticated semi-analytical models
deal with stationary
self-similar investigations gradually extending the~\citet{Blan82}
model with more physical effects, starting from simple vertical mass flux 
prescriptions~\citep{Ward93} to realistic disk equilibria where ambipolar
diffusion \citep{Li96}, resistivity \citep{Ferr97} or both viscosity and
resistivity are adequately incorporated~\citep{Cass00a}. These studies bring 
deep insight in the physical conditions prevailing at the disk surface
required to launch jets, but fail to give realistic jet
topologies in the super-Alfv\'enic region. The latter has spurred a
variety of numerical MHD studies, aiming at a more realistic description of 
the trans-Alfv\'enic flows. The complexity of the dynamics of the 
accretion-ejection flow
has forced many authors to either focus on jet dynamics alone
\citep{Usty95,Ouye97,Kras99,Usty99} or to study disk-outflow dynamics over very
short timescales~\citep{Ushi85,Mats96,Kato02}. The latter class of
  studies were done using ideal MHD framework which is inconsistent with
  long-term jet production since the frozen-in magnetic structure is
  advected with the disk material. This leads to a rapid destabilization of the
  system due to a magnetic flux accumulation in the inner part of the
  disk. Moreover, in most of the previous studies, the accretion disk is
  modeled as a non-Keplerian thick torus without any outer mass inflow
  which would mimic the mass reservoir of the disk outer regions. Here again
  this compromises the long-term jet production since accretion will end
  rapidly because of the lack of disk material.\\ 
In contrast to
these studies, \citet{cass02} (hereafter CK02) recently presented MHD
computations demonstrating the
launching of super-fastmagnetosonic, collimated jets from a 
resistive accretion disk over a large number of dynamical timescales. 
These axisymmetric simulations heavily relied on the general requirements 
identified by self-similar analytical models, in particular the
necessity for equipartition disk regions
together with sufficiently bent magnetic surfaces \citep{Blan82,Ferr95}.
Nevertheless, for simplicity CK02 replaced 
the energy equation by an adiabatic polytropic relation which
did not enable us to explore the energetics of the resulting
accretion-ejection flow. This is of primary interest for realistic
modeling, since accretion disks supporting jets display a disk
luminosity much smaller than expected in the framework of any
currently accepted `standard' model.         

\subsection{Under-luminous accretion disks and hot collimated jets}

The detection of accretion-ejection motions is typically achieved through
multi-wavelength observations. Emissions from the disk and from an associated
outflow differ since they are produced by different mechanisms in
media of widely different densities and temperatures. Accretion disks 
displaying outflows always present low-radiative efficiency, and this is true
in binary stars \citep{Rutt92}, microquasars \citep{Mira98} or AGN
\citep{Dima00,Dima03}. One
scenario to explain the lack of disk emission invokes a 
weak coupling between ions and electrons, resulting in
electron temperatures being much lower than ion temperatures
\citep{Shap76,Rees82}. This scenario 
has led to the concept of ``advection-dominated accretion flows'' (ADAF,
see e.g. \citet{Nara95} and references therein) where energy released 
by accretion of matter is primarily transferred to heat the ions which are
ultimately advected onto the central object. If this is a black hole, 
this energy is then simply lost. Recently, \citet{Blan99} presented an
alternative to ADAF, where inclusion of an outflow from the disk 
was suggested, in which case a part of the accretion energy is sent into 
outflowing hot matter. Although several numerical MHD simulations of
radiatively inefficient accretion disks have now been
reported in the literature (see \citet{Igum03} and references therein),
they have so far been unable to describe self-consistently both 
radiatively inefficient accretion flow and persistent, hot and 
collimated jet outflow.

The aim of the present paper is to compute 
radiatively inefficient magnetized accretion-ejection flow where both
the inward accretion and the upward, transmagnetosonic ejection flow 
can provide a way out for the energy
released by accretion. This is achieved by fully accounting for 
an appropriate energy equation in the MHD description of the
interaction of an accretion disk
with a large scale magnetic field. We present the theoretical and numerical
background of our model in Section~2. In Section~3, we display
results obtained from simulations of a resistive accretion disk with full
energy consideration. In Section~4, we analyze the temporal
evolution of the energy balance including both disk and jet powers. 
We conclude in Section~5 with a summary and open issues to address in
future work.

\section{Magnetized Accretion-Ejection Flow}

We aim to model a magnetized accretion disk launching jets. 
The description of its plasma dynamics is achieved in an MHD
framework. We first present the full set of resistive MHD equations, then
discuss initial and boundary conditions. We conclude this section by
explaining the physical normalization employed in our simulations,
since this makes them applicable to very different systems (YSO to AGN).
   
\subsection{MHD equations}
\label{MHDequa}
The full set of MHD equations expresses conservation of macroscopic
quantities as mass, momentum and energy. The mass conservation is
\be
\frac{\p \rho}{\p t} = -\nab\cdot(\rho\vv),
\label{mhd1}
\ee  
\noindent where $\rho$ is the plasma density and $\vv$ the velocity of the
fluid. Since we are assuming axial symmetry, note that only the \pol
component of velocity is involved in Eq.~(\ref{mhd1}). The momentum
conservation accounts for forces arising from thermal pressure $P$, Lorentz
force, centrifugal force and gravity, namely
\be
\frac{\p\rho\vv}{\p t} + \nab\cdot(\vv\rho\vv-\BB\BB) +
\nab\left(\frac{\BB^2}{2}+P\right) + \rho\nab\Phi_G =0 ,
\label{mhd2}
\ee
\noindent where $\BB$ is the magnetic field and
$\Phi_G=-GM_*/(R^2+Z^2)^{1/2}$ is the Newtonian gravity
potential. Coordinates $(R,Z)$ stand for cartesian coordinates in the \pol
plane. The temporal evolution of the magnetic field is governed by the
induction equation which must take into account the effect of a temporally and
spatially varying resistivity $\eta$, 
\be
\frac{\p \BB}{\p t} + \nab\cdot(\vv\BB - \BB\vv) = -\nab\times(\eta \JJ), 
\label{mhd3}
\ee
\noindent where $\JJ$ is the current density defined by the Maxwell-Amp\'ere
equation 
\be
\JJ = \nab\times \BB \ .
\label{mhd4}
\ee 
We have adopted units where vacuum permeability $\mu_o$ is equal to
unity. The total energy density $e$ is defined as
\be
e = \frac{\BB^2}{2} + \frac{\rho\vv^2}{2} + \frac{P}{\gamma -1} +\rho\Phi_G,
\label{mhd5}
\ee
\noindent where $\gamma=5/3$ is the specific heats ratio of a
non-relativistic plasma. The energy equation in a resistive framework then
reads  
\be 
\frac{\p e}{\p
  t}+\nab\cdot\left[\vv\left(e+P+\frac{B^2}{2}\right)-\BB\BB\cdot\vv \right]= \eta \JJ^2 -\BB\cdot(\nab\times\eta\JJ) .
\label{mhd6}
\ee
\noindent Note
that resistivity, operational in the disk alone, 
provides a local Joule heating. Ohmically heated plasma can then 
be advected by both accreting and ejecting flow in the case of a
radiatively inefficient plasma. The last relation closing the set of
equations links pressure and density to temperature $T=P/\rho$, as we assume 
the plasma to be ideal. 

\subsection{Initial conditions}\label{secic}

The initial configuration of the system is very close to the one adopted
in CK02. The computational domain is $[R,Z]=[0,40]\times[0,80]$ with a
resolution of $104\times 204$ cells and
includes a sink region around
the origin to avoid the gravitational singularity. The size of the sink region
in CK02 was several grid cells long (up to the inner disk radius setting our
unit length) and only two cells
high. If one wants to simulate systems ranging from YSO (Newtonian
gravity) to black holes (with for instance a pseudo-Newtonian 
gravity~\citet{Pacz80}, suitably
describing plasma dynamics from several Schwarzschild radii outwards), we have
to take into account the fact that the gravitational singularity does not
always occur exactly at the origin but can be located at some distance from
it (in the case of a black hole). Therefore, we now modify the size
of the sink region, and design it to cover one unit length both along 
$R$ and $Z$.
For a complete description of the boundary conditions, for both the
treatment of this interior sink region as well as the domain boundaries,
we refer to CK02.

We briefly list our nearly unchanged initial conditions for density and
velocity, namely 
\begin{eqnarray}
\rho(R,Z)&=&
max\left(10^{-6},\frac{R^{3/2}_o}{(R^2+R_o^2)^{3/4}}\right.\nonumber\\
&\times&\left.
max\left(10^{-6},\left[1-\frac{(\gamma-1)Z^2}{2H^2}\right]\right)^{1/(\gamma-1)}\right),  \nonumber\\
V_R(R,Z) &=&
-m_s\frac{R_o^{1/2}}{(R_o^2+R^2)^{1/4}}\exp(-\frac{2Z^2}{H^2}) = V_Z(R,Z)\frac{R}{Z}, \nonumber \\
V_{\theta}(R,Z)&=&(1-\epsilon^2)\frac{R_o^{1/2}}{\epsilon
  (R_o^2+R^2)^{1/4}}\exp(-\frac{2Z^2}{H^2})\ .
\label{Init1}
\end{eqnarray}
\noindent Note that in the present paper we have decreased the initial density
range compared to CK02 for computational performance. However, we still
cover 6 decades in density contrast at $t=0$. Similar to
CK02, we assume the initial thermal pressure to be polytropic such that
$P=\rho^{5/3}$. The disk height $H=\epsilon R$ increases linearly with
radius, with aspect ratio $\epsilon=0.1$ suitable for a thin disk. 
The constant $R_o=4$, while the
parameter $m_s=0.1$ is consistent with initial subsonic accretion
motion. The $m_S$ parameter range is restricted by both subsonic
  accretion motion and the minimal bending of the \pol magnetic surface in
  order to achieve jet production \citep{Blan82}. Indeed the induction
  equation shows that a steady-state magnetic configuration with bent
  magnetic surface above a thin disk requires both resistivity and minimal
  radial motion since 
\begin{equation}
\eta\frac{\p B_R}{\p t} \sim -V_RB_Z \ .
\end{equation}
\noindent The choice of $m_s$ is then to be smaller but close to unity.\\ 
The initial magnetic field configuration is purposely taken very
different from CK02. Where CK02 started from a radially stratified, but
purely vertical field, here we start with an open \pol magnetic
structure to address the issue of jet collimation adequately. 
The main difficulty is to find an initially bent magnetic configuration
which reconciles several constraints like
symmetry at the equatorial disk plane and at the jet axis with the obvious
requirement of $\nab\cdot\BB=0$. Moreover, if one wants to have an almost
constant plasma beta $\beta_P=B^2/P\sim 1$ inside the accretion disk, a 
necessary condition for jet launching, the radial variation of the
magnetic components must be close
to a power-law of index $-5/4$. In order to fulfill the above statements we
adopt the magnetic configuration ($\beta_P=0.6$ in our simulation) which
exploits the function 
\begin{eqnarray}
F(R,Z) &=& \sqrt{\beta_P}\frac{R_o^{5/4}R^2}{(R_o^2+R^2)^{5/8}}\frac{1}{1+\zeta
  Z^2/H^2}, \nonumber \\
\end{eqnarray}
and given in full by
\begin{eqnarray}
B_R(R,Z) &=& -\frac{1}{R}\frac{\p F}{\p Z}, \nonumber\\
B_Z(R,Z) &=& \frac{1}{R}\frac{\p F}{\p R} +
\frac{\sqrt{\beta_P}}{(1+R^2)}, \nonumber \\
B_{\theta}(R,Z) &=& 0\ . 
\label{Init2}
\end{eqnarray}     
\noindent The extra term appearing in the $B_Z$ definition avoids
vanishing vertical magnetic field at the jet axis (this term rapidly vanishes
as $R$ increases) and does not interfere
with the constraint $\nab\cdot\BB=0$. The initial magnetic field
configuration is not force-free but prevents from any initially-induced
collimation since magnetic surfaces tend to widen (outwardly decreasing
magnetic pressure). On figure~(\ref{F1}), we have
displayed this initial \pol magnetic configuration, where we have
set $\zeta=0.04$ ($\zeta$ controlling the initial bending of magnetic
surfaces).
  
The total energy density $e$ is then found from the 
previous quantities according to equation~(\ref{mhd5}). The
anomalous resistivity $\eta$ is prescribed in the same way than in
CK02, namely 
\begin{equation}
\eta=\alpha_m V_A\mid_{Z=0} H \exp\left(-2 \frac{Z^2}{H^2}\right),
\end{equation}
which is an $\alpha$-prescription ($\alpha_m=0.1$) where the Alfv\'en velocity
replaces the sonic speed~\citep{Shak73}. Through the dependence on
the Alfv\'en velocity, this becomes a spatio-temporally varying profile
which essentially vanishes outside the disk.

 \subsection{Numerical code: VAC}\label{s-vac}

The numerical calculations presented here are done using the Versatile
Advection Code (VAC, see \citet{Toth96a} and {\tt
http://www.phys.uu.nl/}$\sim${\tt toth}). We solve the full set of
resistive MHD equations under the assumption of a cylindrical
symmetry. The initial conditions described above are time advanced using
the conservative, second order accurate Total Variation Diminishing
Lax-Friedrichs~\citep{Toth96b} scheme with minmod limiting applied on the
primitive variables. We use a dimensionally unsplit, explicit
predictor-corrector time marching.  To enforce the solenoidal character of
the magnetic field, we apply a projection scheme prior to every time 
step~\citep{brack80}.  

\subsection{Accretion disk properties} 
\label{ADProp}

The initial conditions discussed in section~\ref{secic}
are inspired from self-similar descriptions of accretion disks. 
It is important to realize that through the careful choice of normalization
used in these expressions, our results will be applicable 
to {\em both} YSO and AGN systems. Indeed, units are now expressed in terms
of inner disk radii $R=1$, velocities from $\Omega_K H\mid_{R=1}$, and
density values (very close to or) at $R=1$. We here list what this
dimensionalization implies for both black hole and YSO systems. 

In this particular framework, the
{\it standard} accretion disk model \citep{Shak73,Novi73,Lynd74}
describes astrophysical accretion disks provided that the accretion is
sub-Eddington. If radiative pressure is neglected, it can be shown
that the plasma temperature scales as 
\begin{eqnarray}
T_o &=& \frac{\tilde{\mu} m_p\Omega_K^2 H^2}{k_B}=
10^{11}\left(\frac{R}{R_S}\right)^{-1} K,\nonumber\\
&=& 10^{4}\left(\frac{M_*}{M_{\odot}}\right)\left(\frac{R}{0.1
  AU}\right)^{-1} K .
\label{Set1}
\end{eqnarray}    
\noindent where $\tilde{\mu} m_p$ is the mean molecular mass of the plasma
(proton mass $m_p$), $H=\epsilon R$ is the disk height, 
$\Omega_K=(GM_*/R^3)^{1/2}$ the
Keplerian angular velocity and $k_B$ the Boltzmann's constant. 
The first expression contains $R_S=2GM_*/c^2$, which stands for the
Schwarzschild radius, and is applicable in the case of a black hole. The
second expression is to be used for a YSO disk.

This temperature scaling combined
with the definition of the accretion rate $\dot{M}_a$ leads to all other 
dimensional quantities. When applying our results to a certain system with
inner radius $R$, central object mass $M_*$ and accretion rate
$\dot{M}_a$, the dimensional scale factors are:
\begin{eqnarray}
V_{\theta,o} &=& \epsilon^{-1}V_{R,o} = \Omega_KR = 2.1\times
10^{10}\left(\frac{R}{R_S}\right)^{-1/2}\ cm/s\nonumber\\
&=& 9.5\times 10^6 \left(\frac{M_*}{M_{\odot}}\right)^{1/2}\left(\frac{R}{0.1
  AU}\right)^{-1/2}\ cm/s\nonumber \\
\rho_o &=& \frac{\dot{M}_a}{4\pi V_{R,o}RH}= 2.7\times
  10^{5}\left(\frac{\dot{M}_a}{M_{\odot}/yr}\right)\left(\frac{M_*}{M_{\odot}}\right)^{-2}\nonumber\\
&\times&\left(\frac{R}{R_S}\right)^{-3/2}\ g/cm^3\nonumber\\
&=& 2.4\times
  10^{-12}\left(\frac{\dot{M}_a}{10^{-7}M_{\odot}/yr}\right)\left(\frac{M_*}{M_{\odot}}\right)^{-1/2}\nonumber\\
&\times&\left(\frac{R}{0.1AU}\right)^{-3/2} \ g/cm^{3}\nonumber\\
P_o &=& \rho_o\Omega_K^2H^2 = 1.2\times
10^{24}\left(\frac{\dot{M}_a}{M_{\odot}/yr}\right)\left(\frac{M_*}{M_{\odot}}\right)^{-2}\nonumber\\
&\times& \left(\frac{R}{R_S}\right)^{-5/2}\
erg/cm^3 \nonumber\\
&=&
2.1\left(\frac{\dot{M}_a}{10^{-7}M_{\odot}/yr}\right)\left(\frac{M_*}{M_{\odot}}\right)^{1/2}\nonumber\\
&\times& \left(\frac{R}{0.1AU}\right)^{-5/2}
erg/cm^{3}\ . 
\label{Set2}
\end{eqnarray}
\noindent Again, the first expressions are to be used for black hole systems,
while the second lines are for YSO.

The last quantity is the magnetic field that we set to be close to
equipartition with thermal pressure, a necessary condition for jet
launching. Indeed, the above temperature statement implies that
  thermal pressure naturally overcomes the gravitational pinching in
  standard accretion disk ($P\sim \rho\Omega_K^2H^2$). Magnetized 
  disks are also prone to magnetic pinching so a natural equilibrium
  of the disk is such that $P \geq B^2/\mu_o$. The former constraint coupled
  with 
  the MHD Poynting flux amplitude threshold to produce
  super-fastmagnetosonic, collimated outflows leads to a disk configuration
  where $P\sim B^2/\mu_o$ \citep{Ferr95}. The magnetic field will then be 
\begin{eqnarray}
B_o &=& 3.9\times
10^{12}\beta_P^{1/2}\left(\frac{\dot{M}_a}{M_{\odot}/yr}\right)^{1/2}\left(\frac{M_*}{M_{\odot}}\right)^{-1}\nonumber\\
&\times&\left(\frac{R}{R_S}\right)^{-5/4}
Gauss\nonumber \\
&=& 7 \beta_P^{1/2}
\left(\frac{\dot{M}_a}{10^{-7}M_{\odot}/yr}\right)^{1/2}\left(\frac{M_*}{M_{\odot}}\right)^{1/4}\nonumber\\
&\times&\left(\frac{R}{0.1AU}\right)^{-5/4}   Gauss .
\label{Set3}
\end{eqnarray}
\noindent Observations indicate that the inner parts of AGN and YSO disks
mainly differ in their temperature (AGN $\geq 10^{9}K$, YSO $\leq
10^{4}K$). This difference is also evident in 
other properties (pressure, magnetic
field), but its main consequence is that these systems have very different
dominant radiative mechanisms operational, since the opacity regime varies 
from fully to partially ionized plasma.
          
\section{Accretion disks and ``hot'' jets} 

In this section, we present the MHD simulations
resulting from the initial configuration presented in the previous section. 
We first
describe the temporal evolution of the accretion-ejection structure, then
display the full jet launching mechanism and finally check the radial force
balance that is reached in the jet. 

\begin{figure*}[t]
\plotone{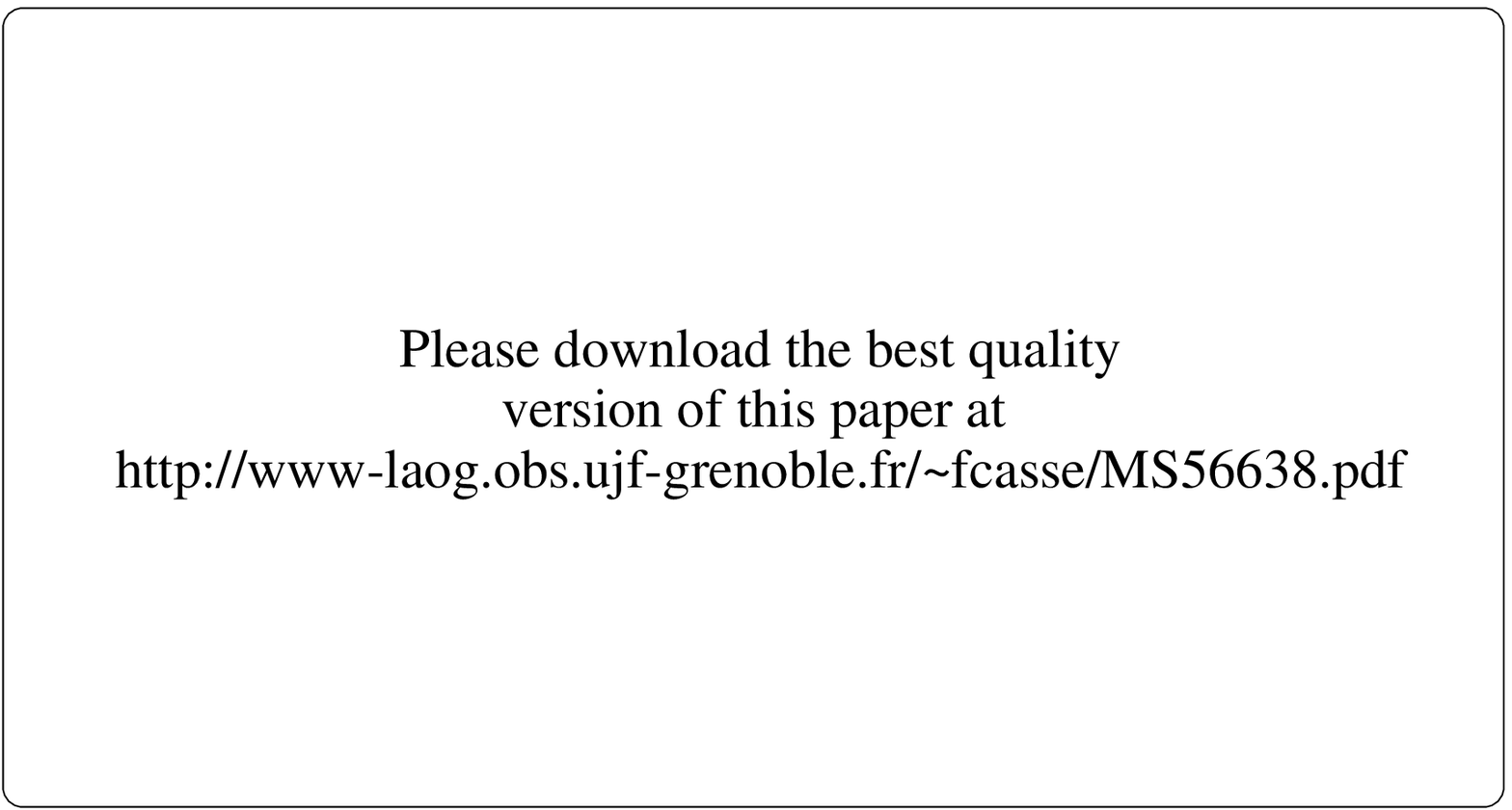}
\caption{Temporal evolution of a resistive accretion disk threaded by a
  \pol magnetic field. Color levels represent
  density level while solid lines stand for \pol magnetic field lines. The time
  unit labeling each snapshot is the number of rotations of the inner
  radius. After a few rotations, outflows are
  escaping from the disk and remain focused despite an initial bent \pol
  magnetic configuration. Once jets are launched, the structure 
  varies little during the remainder of the whole computation.}    
\label{F1}
\end{figure*}
\begin{figure*}
\plotone{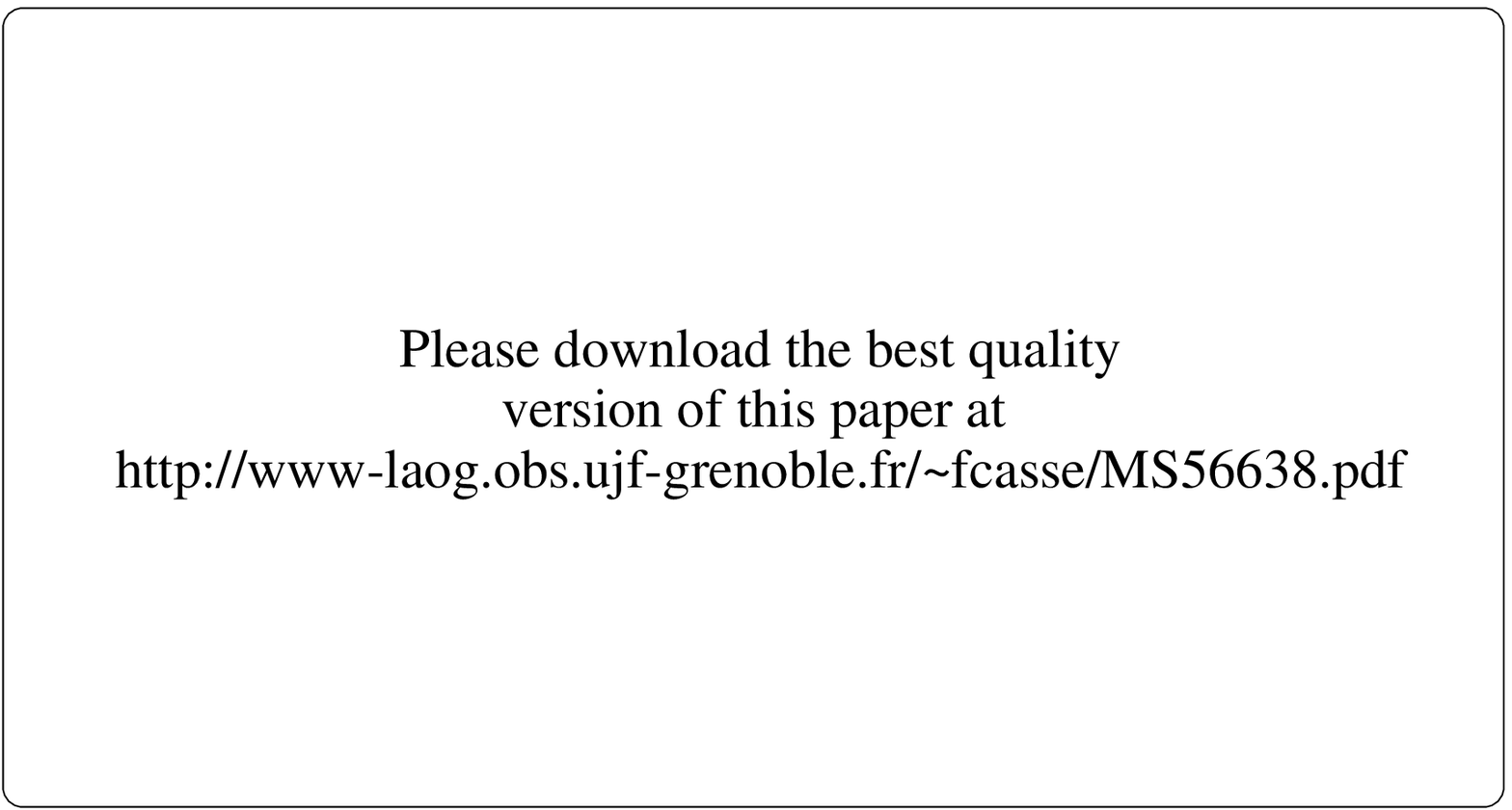}
\caption{Three-dimensional impression of the final near-stationary endstate
 reached in our simulations. Color levels represent a volume rendering of
 plasma density and translucent surface stands for a magnetic
 surface anchored at $R=3$ in the disk. Yellow and blue lines stand for
 magnetic field lines and flow streamline respectively. As can be seen,
 magnetic field lines are twisted by disk rotation which provokes mass
 acceleration as shown by the flow streamline that is initially accreting
 toward the central object and then turns into the jet.}
\label{F1B}
\end{figure*}

\label{HotJ}
\subsection{Temporal evolution of the flow}
\label{TempE}
The result of numerically simulating the full MHD dynamics of the resistive
magnetized accretion disk is displayed in Fig.~\ref{F1}.  The six snapshots
represent six \pol cross-sections of the structure at different stages of
its temporal evolution. The initial accretion disk configuration is close
to a radial equilibrium where Keplerian rotation balances gravity of the
central mass. The rotation of matter twists the initially purely \pol
magnetic field lines such  that outflows appear at the surface of the
disk. The outflow is continuously emitted during the further evolution of
the system. As seen in the density levels of the  various snapshots, this
outflow is less dense ${\cal O}(10^{-3})$ than the accretion disk  ${\cal
O}(1)$ but denser than the surrounding medium ${\cal O}(10^{-6})$ by
several orders of magnitude.  The evolution of both the mass outflow and
the magnetic field topology gets closer to an equilibrium as time increases
since the \pol velocity component of the outflow becomes parallel to the
\pol magnetic field. Moreover, the outflow is radially collimated and
crosses the Alfv\'en and fast magnetosonic speeds, to become
superfastmagnetosonic well before it reaches the top boundary of the
computational domain (see also CK02).  It can then safely be qualified as a
jet. In Fig.~\ref{F1B}, we present  a three-dimensional representation of
the final stationary endstate reached. We visualized a magnetic surface
which is anchored at $R=3$ which nicely shows its initial fanning out as
well as its eventual collimation. Selected fieldlines are shown in yellow,
revealing their helical character. We also plot a streamline (in blue)
which shows the path of a plasma parcel that is initially accreting within
the disk in a spiral fashion, until it  reaches the jet launch area where
it gets propelled into the jet. Finally, the figure also contains a volume
rendering of the density in accretion disk-jet system, which shows that the
density structure at the foot of the jet is in fact convex  since full
collimation is only achieved from some vertical distance above the disk. In
our simulations full flow collimation occurs beyond $Z=50$, which corresponds
to typically several tens of disk scale heights.

\begin{figure}
\plotone{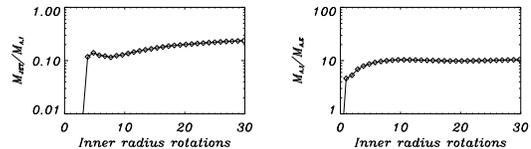}
\caption{Ejection mass rates and inner accretion rates as a function of
  time. The inner accretion rate, normalized to the constant value of the outer
  accretion rate, strongly increases
  during the very early stages and then remains almost constant during the rest
  of the simulation. The ejection rate increases slowly with time,
  and reaches a final value of the order of $20\%$ of the inner accretion rate.}
\label{F1bis}
\end{figure}

In order to quantify the temporal evolution of both accretion and ejection
flows, we can have a look at both accretion and ejection mass rates. These
rates also trace the efficiency of the magnetic field to remove angular
momentum from the disk (to allow accretion) and to accelerate matter in the
jet (which creates a vertical mass flux as seen on Fig.~\ref{F1}). On
Fig.~\ref{F1bis}, we have displayed these mass rates as a function of time,
$\dot{M}_{A,I}$ and $\dot{M}_{A,E}$ being accretion rates measured at inner
$R_I=1$ and external radius $R_E=40$, respectively,
while $\dot{M}_{JET}=2\pi\int_{0}^{R_E}R\rho V_ZdR$ is
the ejection mass rate measured at a given $Z$ above the disk surface. 
On Fig.~\ref{F1bis}, the inner accretion rate is normalized with respect to
the value of the outer accretion rate which by means of our
boundary conditions (fixing the mass flux) is set to be constant. 
The strong increase of the inner accretion rate with time is a
clear translation of the role of the magnetic torque in the disk: as the
simulation begins, the twist of magnetic field lines provoked by disk
rotation creates a \tor component for the magnetic field which directly leads
to a braking magnetic torque. The azimuthal
braking of matter reduces the centrifugal
force and thus allows matter to be attracted by gravity at a higher
rate. Nevertheless, as seen on Fig.~\ref{F1bis}, the increase rapidly
disappears to let the inner accretion rate settle to an almost constant
value. The existence of this plateau accounts for an accretion disk where
matter rotation and magnetic tension reach an equilibrium state. The
ejection rate is slowly rising and reaches a final value which is of
the order of $ 22\%$ of the inner accretion rate. During the final stages
both ejection and accretion rates remain almost constant and enable us to
study in detail the accretion-ejection connection. The final ratio of inner
to outer accretion rates is of the order of ten. This value larger than
unity arises from our choice to impose a constant outer accretion rate
during the simulation, thus not enabling the structure to adjust itself to
the outer part of the accretion disks that lie outside the
computational domain. 

Note that these accretion rates have already
settled on constant values after about 10 inner orbital periods. We have 
evolved the system further to 30 inner orbital periods, and as seen in the
snapshots from Fig.~\ref{F1}, the overall structure remains rather
stationary during that time. As the jet is confined to within 20 inner
radii, the final time $T=30$ covers $1.3$ rotation period as measured at the
external foot of the jet, or only $0.13$ rotation period of the outer radius at $R=40$.
However, as will become clear from the further analysis of the force
balances achieved in the inner disk regions, this is a truly stationary
configuration. Note also that due to the large density contrasts, this 
corresponds to several tens of thousands of CFL limited time steps, which
make our simulations computationally challenging. 

Note that our endstate is in contrast to recent 3D simulations of 
radiatively inefficient accretion flows~\citep{Igum03}, 
whose authors claim that we are focusing on a transient phase of the flow.
However, the quasi-steady endstate observed in~\citet{Igum03} is
one of dominant magnetic field in the entire computational domain.
This must be due to the imposed continuous injection of \pol
magnetic flux adopted in that study. In contrast, our simulations do not
assume an infinite reservoir of magnetic flux, and the magnetic field
settles into a stationary state where inward advection is balanced by
outward slippage due to the disk resistivity. Moreover, in our simulation
the level of magnetic pressure is always close to unity either in the
accretion disk or in the jet.

\subsection{Force balance and current circuit}
\label{Curr}

\begin{figure*}[t]
\plotone{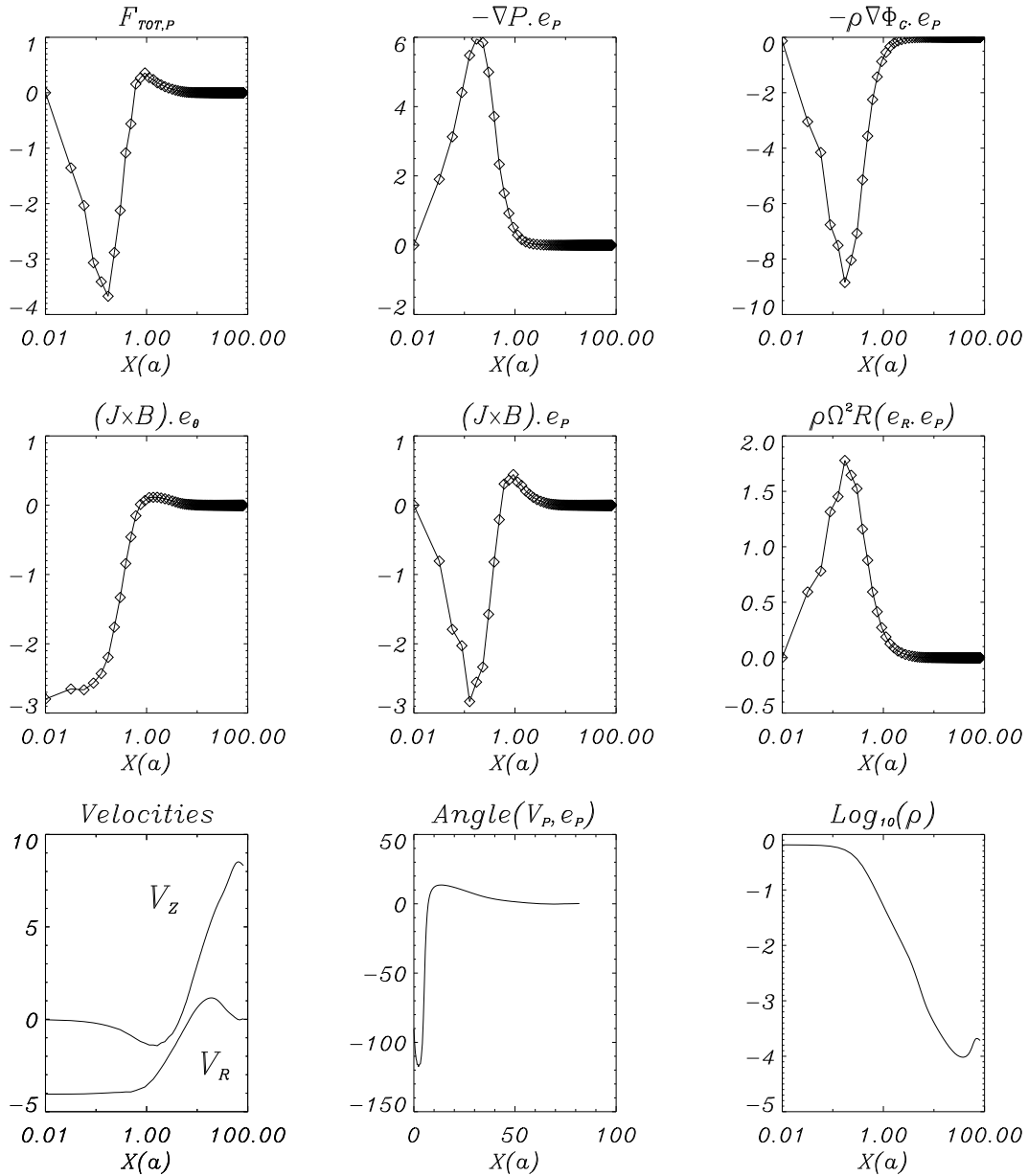}
\caption{Plots of forces projected along a given magnetic
  surface as a function of the curvilinear abscissa along this magnetic
  surface $X(a)$ at T=30. The footpoint of this surface is at $R=3$. From
  left to right and from top to bottom: total force,
  thermal pressure gradient, gravity, magnetic torque, \pol magnetic force, 
  centrifugal force, radial and vertical 
  velocities, angle between \pol velocity and \pol magnetic field and
  density. The key to achieve an accretion-ejection
  configuration is to have a magnetic configuration where the Lorentz force
  pinches the disk and accelerates matter in the jet.}
\label{F2}
\end{figure*}

The accretion-ejection zone, as already shown in CK02, is the most
crucial area when simulating the launching of jets. A subtle balance between
different forces must be obtained in order to achieve a continuous emission
of matter, as well as the collimation of the resulting outflow. As an
example, we display in Fig.~\ref{F2} the different forces in the
disk equilibrium at the final time of the simulation from Fig.~\ref{F1}. 
In this figure, the \pol forces are represented along a
given magnetic surface already represented in Fig.~\ref{F1B}
as well as the magnetic torque, \pol velocity and
temperature. The vector $\ep$ occurring in Fig.~\ref{F2} is defined as
$\BB_P/|\BB_P|$. The upper-left plot represents the total force applied on
the plasma which is negative in the disk and becomes positive near the disk
surface. As already shown in CK02, this configuration enables the matter
below the disk surface to be pinched and to remain in an accretion regime
while beyond the disk surface, matter is accelerated along magnetic field 
lines,
leading to a jet. This balance is obtained from competing forces where
thermal pressure gradient and centrifugal force counteract magnetic
and gravitational pinching. The crucial point of the accretion-ejection
connection is to accurately capture the sign change of the sum of all forces
in the numerical stationary state. This
change of sign can only be achieved if the \pol magnetic force along the 
magnetic
surface changes its sign near the disk surface, as shown on Fig.~\ref{F2}. Since
this force is directly related to the magnetic torque, 
\be
(\JJ\times\BB)\cdot\BB_P = -(\JJ\times\BB)\cdot\BB_{\theta},
\ee              
\noindent this also implies that the magnetic torque changes its sign there, as
also shown in Fig.~\ref{F2}. This configuration is at the core of
magneto-centrifugal acceleration since \tor acceleration (or braking)
determines \pol acceleration (or braking) of matter.
The \pol velocity field shown in Fig.~\ref{F2} confirms the above statements,
since this field is
consistent with an accretion motion in the disk (dominant negative radial
velocity) while the vertical velocity becomes dominant beyond the disk
surface. The angle between the \pol velocity and \pol magnetic field is
displayed on the bottom middle panel and again confirms the required
accretion-ejection configuration where flow is perpendicular to the
magnetic field in the disk and becomes essentially parallel to it in the jet
region. 

The plasma temperature levels are displayed in the left panel of Fig.~\ref{F6} and exhibits an
interesting behavior. In CK02, we were using a polytropic relation
which was constraining the temperature to be proportional to $\rho^{\gamma
  -1}$. This 
invariably led to a jet where the temperature is very low because of the
relatively low densities in the jet. As we now include the full
energy equation (and include Ohmic heating), this is no longer happening.

We can see on Fig.~\ref{F6} that the disk
temperature, which was initially smaller than unity at this location, has
increased reflecting the locally operative Joule heating and gas
dynamics. Rewriting energy equation (\ref{mhd6}) by subtracting momentum and
induction equations leads to the temperature equation that reads
\begin{equation}
\frac{1}{\gamma -1}\left(\frac{\p T}{\p t}+\vv_p\cdot\nabla T\right) = \frac{\eta\JJ^2}{\rho} -
T\nabla\cdot\vv \ .
\label{Tequa}
\end{equation} 
\noindent where both Ohmic heating and flow dynamics affect the behavior
of the temperature. Ohmic heating is occurring inside the accretion disk and
inevitably leads to a higher disk temperature than in the initial
stage. In the accretion disk, the velocity divergence is negative since 
$\nabla\cdot\vv=\p V_Z/\p Z + \p V_R/\p R +V_R/R$ where $V_R$  is a slowly
radially varying negative component. This term will also account for a
local heating of the plasma, even in the case where no Ohmic heating would
occur. Assuming the structure to be close to a steady-state, one has
\be 
\vv_p\cdot\nabla T \simeq (\frac{\eta\JJ^2}{\rho} - T\nabla\cdot\vv)(\gamma -1) > 0
\ee   
which is consistent with Fig.~\ref{F6} where isocontours of temperature are
roughly perpendicular to the streamlines in the disk. In the jet region,
temperature isocontours coincide with streamlines as soon as streamlines
have turned from an accretion motion ($V_R<0$) to an ejection motion
($V_R>0$). It is noteworthy that at this precise location we have strong
negative $\nabla\cdot\vv$ which leads to a sudden temperature increase. The
resulting ``hot'' jet is a general property of this kind of outflow since
it does not require local Ohmic heating in the corona. We mention that
performing the same simulation presented here but neglecting Ohmic heating
in the energy equation also gives birth to jets which have similar
temperature than the disk,although slightly lower than
with Ohmic heating included. The relatively small influence of the Ohmic
heating can be explained in our case by the fact that the chosen resistivity is
very low since $\eta=\alpha_m H V_A \leq 10^{-2} R$.\\  

\begin{figure}[t]
\plotone{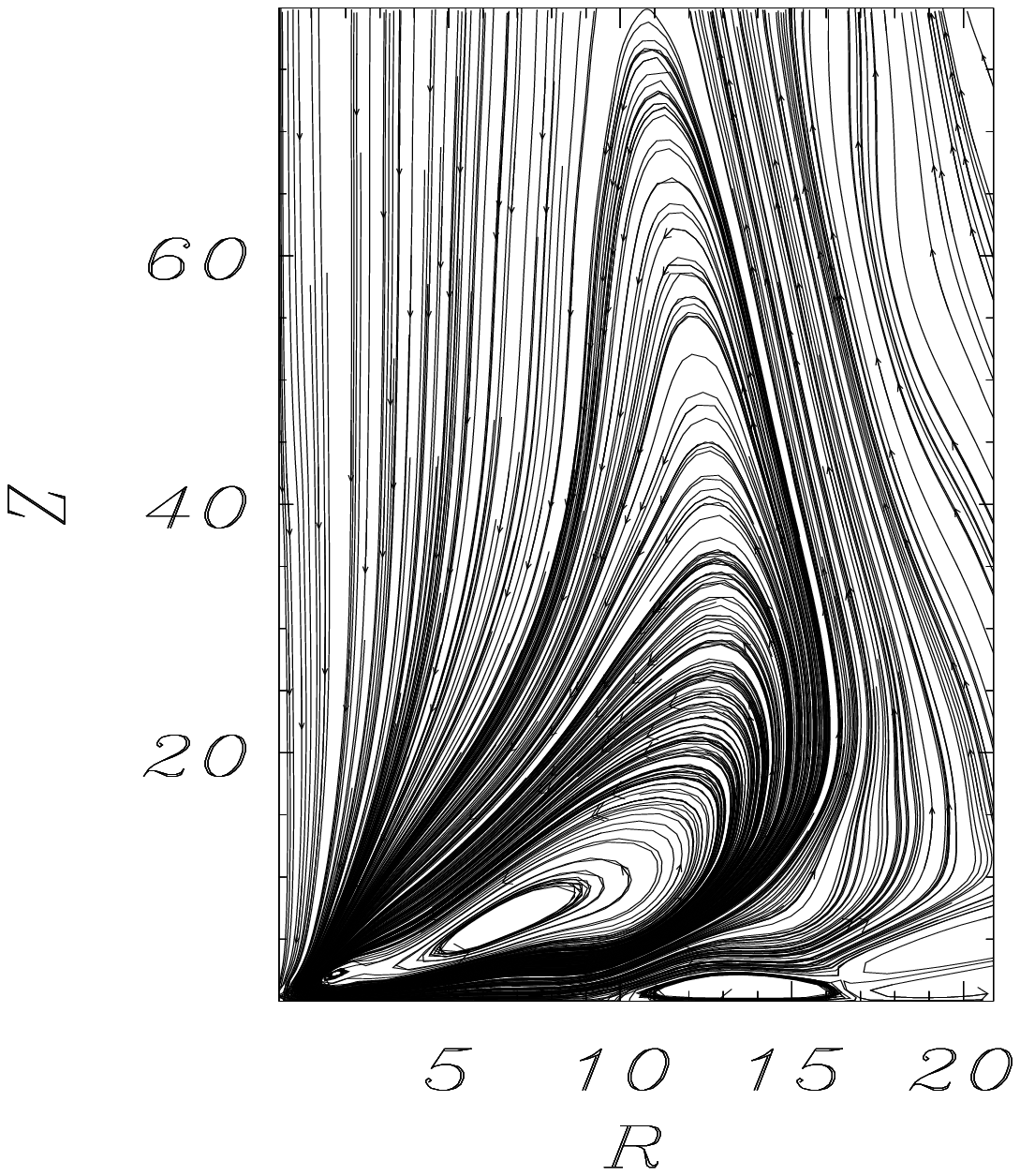}
\caption{Isocontours of current $I\propto RB_{\theta}$. The closed current
  loops are 
  direct evidence of the existence of an accretion-ejection structure where
  magnetic field pinches matter inside the disk and accelerates it beyond
  the disk surface (see Section \ref{Curr}). This circuit is one `wing'
 of a full `butterfly' configuration.}
\label{F2bis}
\end{figure}

It is possible to clarify another aspect of the accretion-ejection mechanism
by analyzing the current circuit established in the final structure. Indeed,
in an axisymmetric
framework, it is easy to calculate the current $I$ at any location thanks
to the Maxwell-Amp\'ere law stating that
\be
I = \int\int\JJ\cdot d{\bf S}=\oint\BB\cdot d{\bf l}=2\pi RB_{\theta} < 0,
\ee
\noindent where the contour in the last integral is a circle of radius $R$
with its center located at the jet axis. Rewriting the \pol magnetic force as 
\be
(\JJ\times\BB)\cdot\ep=-B_{\theta}\frac{\nab I}{2\pi R}\cdot\ep,
\ee
\noindent it is clear that this force is entirely controlled by the local 
variation of the current along the magnetic field line. Looking back at
Fig.~\ref{F2} where the magnetic force projected along the magnetic field
line is represented, we see that the above mentioned change of sign of the
force has a direct implication for the current, namely that the derivative
of the current along the magnetic field line also changes its sign. Looking
now at Fig.~\ref{F2bis}, where isocontours of $I$ are displayed, we can see
that this leads to the formation of closed current loops since the current
$I$ start to decrease and then increase. This kind of current circuit is
representative for a working accretion-ejection structure where both magnetic
braking in the disk and magnetic acceleration in the jet occur. In a full
meridional cross-section through the disk-jet system, this is reminiscent
of a `butterfly' current circuit \citep{Ferr97}.        

\subsection{Jet collimation}

\begin{figure*}[t]
\plotone{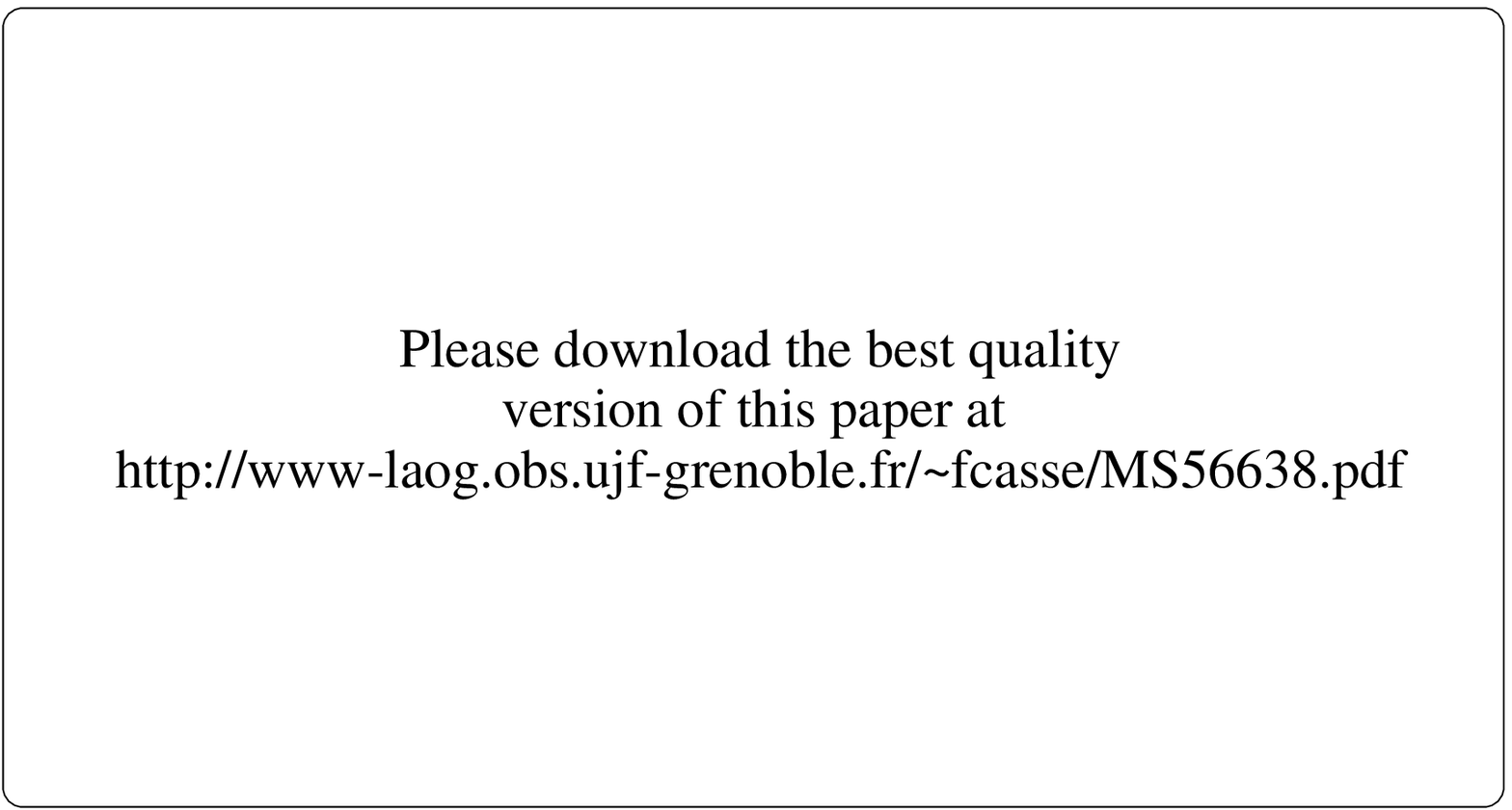}
\caption{Same simulation than the last snapshot of Fig.\ref{F1} but with a square simulation domain and ``force-free'' boundary for the \tor magnetic field
  ($\JJ_p\times\BB_p=0$). The collimation of the jet is still occurring and
  the jet radius is the same than in the previous calculation. The
  dotted-dashed line is the contour of the fast-magnetosonic surface where
  the \pol velocity of the flow equals the local fast-magnetosonic speed.}
\label{F1b}
\end{figure*}

\begin{figure*}[ht]
\plotone{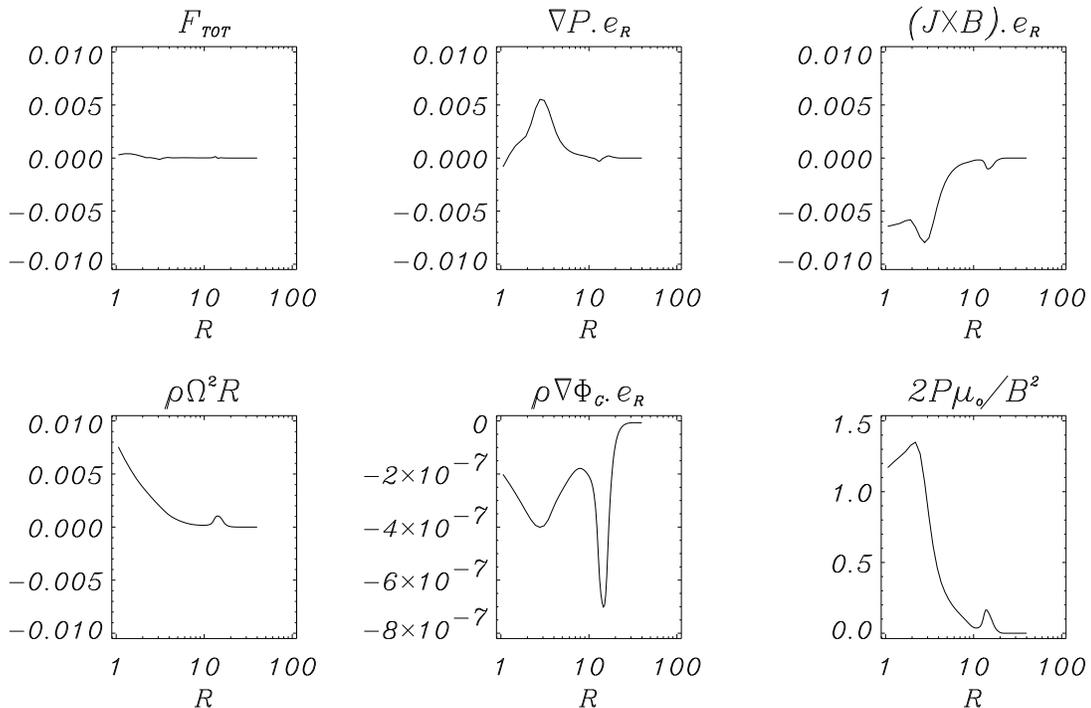}
\caption{Plots of radial forces acting on the jet at Z=60 and at
  T=30. The total force is very small and ensures a good collimation for the
  jet. The radial equilibrium of the jet is achieved thanks to a thermal
  pressure 
  gradient and centrifugal force balancing the radial magnetic force. This
  equilibrium where thermal pressure is of the same order than magnetic
  pressure is the signature of a ``hot'' jet. }
\label{F3}
\end{figure*}

In CK02, we presented MHD simulations of accretion disks
launching jets where the resulting outflow was collimated. In that work,
we were able to show that the radial force balance in the jet was achieved 
by magnetic tension counteracting magnetic pressure effects (a cold jet). 
Nevertheless, it was not a priori obvious
that this collimation is not affected by our choice of an initial
purely vertical magnetic field configuration. In the present paper, we
purposely 
choose a different initial magnetic configuration where the \pol surfaces are 
strongly bent at $T=0$ as shown in Fig.~\ref{F1}. Still, the resulting jet 
is well collimated, which is a direct proof that the collimation of this 
flow is arising self-consistently. We display in Fig.~\ref{F3} 
the radial balance of the jet for the same
snapshot than in Fig.~\ref{F2} at $Z=60$. Clearly, the total force is 
vanishingly small, proving that there is equilibrium in the jet, 
but the situation is quite different than in
CK02. Indeed, now the thermal pressure is of the same order than
the magnetic pressure which leads to a qualitatively different radial jet
equilibrium. The thermal pressure gradient as well as the centrifugal force
are now balancing the total radial magnetic force. This kind of force
balance between thermal pressure and magnetic force is another clue to the
occurrence of a ``hot'' jet where thermal energy is of the same order than
the magnetic energy.\\
 In order to rule out any artificial collimation of the jet, we have
performed the same simulation with different box size and boundary
conditions for the magnetic field. Since \citet{Usty99}, it has been shown
that axisymmetric magnetic field having open boundaries may induce
artificial collimation forces due to imposed vanishing radial current
in the ghost cells. The authors showed that one way to obtain accurate
numerical MHD flow solutions is to switch from open boundary on the \tor
magnetic field to a ``force-free'' one (where $B_p\cdot\nabla(RB_{\theta}=0$)
and to perform the simulation on a square domain in $R$ and $Z$. We have
followed exactly the same approach than \citet{Usty99} and obtained a
solution very close to the one presented in Fig.\ref{F1}. For instance, we
present in Fig.\ref{F1b} the density and \pol magnetic field structure as
well as \pol streamlines of the flow at $T=30$. The collimation is
undoubtedly occurring since the  jet magnetic and velocity field are completely
vertical and super-fastmagnetosonic as the flow is escaping from the
domain. The fast-magnetosonic surface is shown on Fig.~\ref{F1b} with a
dotted-dashed line and stands for the location where \pol velocity equals
the local fast-magnetosonic velocity. 
    
\section{Radiatively inefficient accretion disk launching jets}
\label{EnerBud}
The inclusion of the energy equation in the MHD description
enables us to investigate the energetics of the magnetized
accretion-ejection structure. For instance, in
classical hydrodynamical ADAF, it is shown that the energy released by
accretion of matter is balanced by creation of entropy within the disk. In
the present simulations, the occurrence of a dynamically significant
magnetic field opens new ways to transfer and to use accretion energy. We
will begin this section by defining the different powers relevant for the
disk energetics. Subsequently, we will analyze the temporal evolution
of these quantities in order to identify the predominant
way of energy transport in the flow.

\subsection{MHD Energy equation} 
In Section \ref{MHDequa}, we presented the MHD equations and in
particular the energy equation~(\ref{mhd6}). This equation can equivalently be
rewritten in a full conservative form as 
\be
\frac{\p e}{\p
  t}+\nab\cdot\left[\vv\left(e+P+\frac{B^2}{2}\right)-\BB\BB\cdot\vv+\eta\JJ\times\BB \right]= 0 \ .
\label{EC1}
\ee
Using Ohm's law, we can replace the current density $\JJ$ by
the electromotive field $\EE=\BB\times\vv + \eta\JJ$ and write the general
energy conservation in a form which is independent of the MHD regime (either
ideal or resistive), 
\begin{eqnarray}
\frac{\p e}{\p
  t}&+&\nab\cdot\left[\rho\vv\left(\frac{v^2}{2}+\frac{\gamma}{\gamma-1}\frac{P}{\rho}+\Phi_G\right)+\EE\times\BB
  \right]= 0  \nonumber \\
\Leftrightarrow \frac{\p e}{\p
  t}&+&\nab\cdot\left[\FF_{acc} + \FF_{MHD}\right] = 0 \ .
\label{EC2}
\end{eqnarray}
The last term is the MHD Poynting flux. In an axisymmetric framework the 
two fluxes that control the
temporal evolution of total energy density are (only the \pol components of
these fluxes are relevant):  
\begin{itemize}
\item{{\it Accretion energy flux $\FF_{acc}$:} the 
dominant component in the disk is the 
radial one since the flow components in the accretion disk obey 
$V_{\theta} \gg |V_R|
  \gg |V_Z|$. In 
standard accretion disk models \citep{Shak73,Novi73,Lynd74}, entropy is
negligible inside the disk so that this flux can be reduced to   
  $\vec{F}_{acc}\sim \rho\vv_R\Phi_G/2$. Note that in the case of a jet,
  the energy flux $\FF_{acc}$ through the disk surface can become significant.}
\item{{\it MHD Poynting flux:} represents the energy carried by
  the magnetic field via angular momentum stored in the magnetic
  field. At the disk surface, where the ideal MHD regime prevails, this flux
  can be written as 
  \be
  \FF_{MHD}=\vv B^2-\BB(\BB\cdot\vv)\simeq \vv B^2+\BB |v_{\theta} B_{\theta}| 
  \label{EBmhd}
  \ee
   which shows that the energy will escape from the disk along flow
   streamlines and magnetic field lines (which are both directed upwards in the
   corona).}     
\end{itemize}
For our MHD simulations presented in the previous section, we can define and
follow the evolution of the following different powers. The energy liberated
by accretion is defined as
\begin{eqnarray}
P_{LIB} &=& P_{MEC}+ P_{ENT}+P_{MHD}, \nonumber
\end{eqnarray}
where 
\begin{eqnarray}
P_{MEC} &=&
-\int\int_{S_I}{\bf dS}_I\cdot\rho\vv\left(\frac{v^2}{2}+\Phi_G\right)\nonumber\\
&-&\int\int_{S_E}{\bf
  dS}_E\cdot\rho\vv\left(\frac{v^2}{2}+\Phi_G\right),\nonumber\\
P_{ENT} &=&
-\int\int_{S_I}{\bf
  dS}_I\cdot\rho\vv\left(\frac{\gamma}{\gamma-1}\frac{P}{\rho}\right)\nonumber \\ 
&-&\int\int_{S_E}{\bf dS}_E\cdot\rho\vv\left(\frac{\gamma}{\gamma-1}\frac{P}{\rho}\right),
 \\
P_{MHD} &=& -\int\int_{S_I}{\bf
  dS}_I\cdot\left(\EE\times\BB\right)-\int\int_{S_E}{\bf dS}_E\cdot\left(\EE\times\BB\right)\nonumber,
\label{Pow1}
\end{eqnarray} 
\begin{figure*}[t]
\plotone{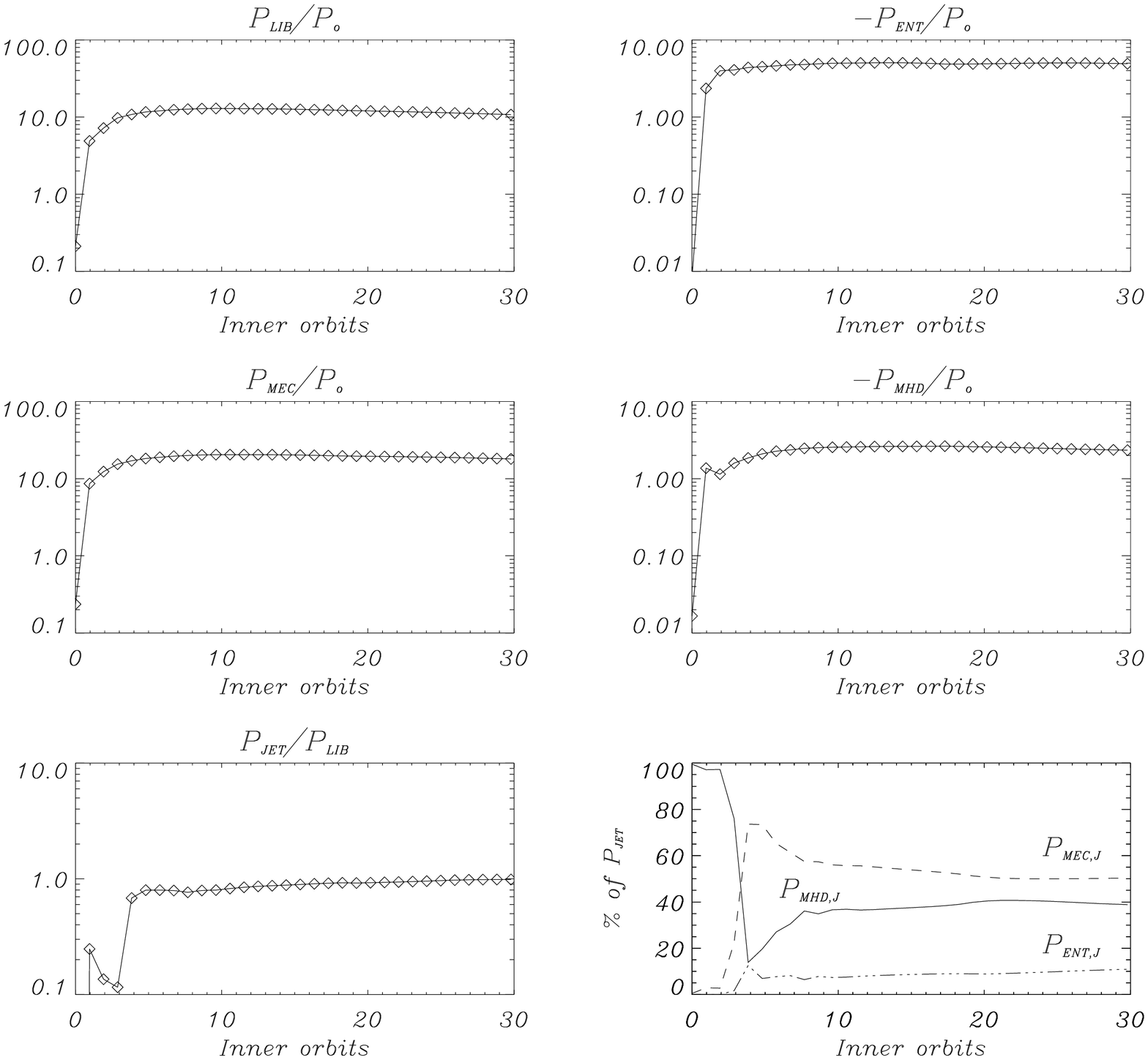}
\caption{Temporal evolution of powers defining the complete
energy budget as presented in Section~\ref{EnerBud}. The
accretion liberated power is non-vanishing in our structure unlike
ADAF-type flow. Indeed, disk enthalpy and magnetic energy fluxes are small
compared to mechanical power which is mainly sent into the jet. This
structure is completely consistent with an under-luminous disk launching bright jets.}
\label{F5}
\end{figure*}
\noindent where the surfaces $S_I$ and $S_E$ correspond to the inner and 
outer vertical cut through the accretion disk (${\bf dS}_I=-2\pi
R_IdZ\er$ 
with $-H< Z <H$  and ${\bf dS}_E=2\pi R_EdZ\er$ with $-H< Z <H$). In
standard disk 
models, the azimuthal velocity of matter is close to Keplerian so that the
energy 
liberated by accretion is $P_{LIB}\sim GM_*\dot{M}_{A,I}/2R_I$ where
$\dot{M}_{a,i}$ is the accretion rate evaluated at the inner radius.
In ADAF models, the enthalpy contribution $P_{ENT}$ is not negligible 
and since it is negative, the liberated
energy (responsible for the disk luminosity in that particular case) is
reduced. The contribution of the radial Poynting flux is more difficult to
estimate inside the disk but we will show that 
it is a small contribution to the
liberated energy.

In the present study, the occurrence of a magnetized jet outflow requires
that we also evaluate the energy flux through the disk corona. Again we define
a jet power $P_{JET}$ such that
\begin{eqnarray}
P_{JET} &=& P_{MEC,J}+ P_{ENT,J}+P_{MHD,J} \nonumber\\
P_{MEC,J} &=& \int\int_{S_{surf}}{\bf dS}_{surf}\cdot\rho\vv\left(\frac{v^2}{2}+\Phi_G\right)\nonumber\\
P_{ENT,J} &=& \int\int_{S_{surf}}{\bf
  dS}_{surf}\cdot\rho\vv\left(\frac{\gamma}{\gamma-1}\frac{P}{\rho}\right)\nonumber\\
P_{MHD,J} &=& \int\int_{S_{surf}}{\bf dS}_{surf}\cdot\left(\EE\times\BB\right)
\label{Pow2}
\end{eqnarray}    
\noindent where ${\bf dS}_{surf}=2\pi RdR\ez$ with $R_I< R<R_E$. In the
disk corona, the vertical velocity of matter can be large due to magnetic
acceleration and will produce a significant vertical energy flux. If the
structure reaches a near-steady state, our energy balance
would read $P_{MEC}+P_{ENT}+P_{MHD}= P_{JET}$. If
neither a magnetic field nor a jet is present, this conservation gives the
usual ADAF-like relation $P_{MEC}\simeq -P_{ENT}$ so that $P_{LIB} \simeq
0$ and the disk will thus be very under-luminous.

\subsection{Under-luminous disks producing jets}

We display in Fig.~\ref{F5} the temporal evolution of all
previously defined powers that characterize the
energetics of the structure. First of all, the upper left plot represents
the temporal evolution of the energy liberated by the accretion of matter
$P_{LIB}$, this power being normalized to the ``standard'' accretion power
expected for a thin disk $P_o= GM_*\dot{M}_{Ao}/2R_I$ where $\dot{M}_{Ao}$ is
the inner accretion rate at $T=0$. This power rapidly increases in early
stages of the evolution and then reaches a plateau where it remains almost
constant at about 10 time its standard value (mainly controled by the inner
accretion rate, see discussion in section~\ref{TempE}). In the three 
following plots, we see the
relative amplitude of each contribution to this power. The mechanical power
$P_{MEC}$ is the dominant contribution while advected enthalpy $P_{ENT}$ and 
Poynting flux $P_{MHD}$ are small and of the same order, with ratios as
\be
\frac{-P_{ENT}}{P_{MEC}} \simeq 0.27 \ , \frac{-P_{MHD}}{P_{MEC}} \simeq
0.1\ .
\ee   
\noindent Note that these two contributions are negative, meaning
that they both advect a portion of the energy
through the inner radius and reduce the available disk power.

\begin{figure*}
\plotone{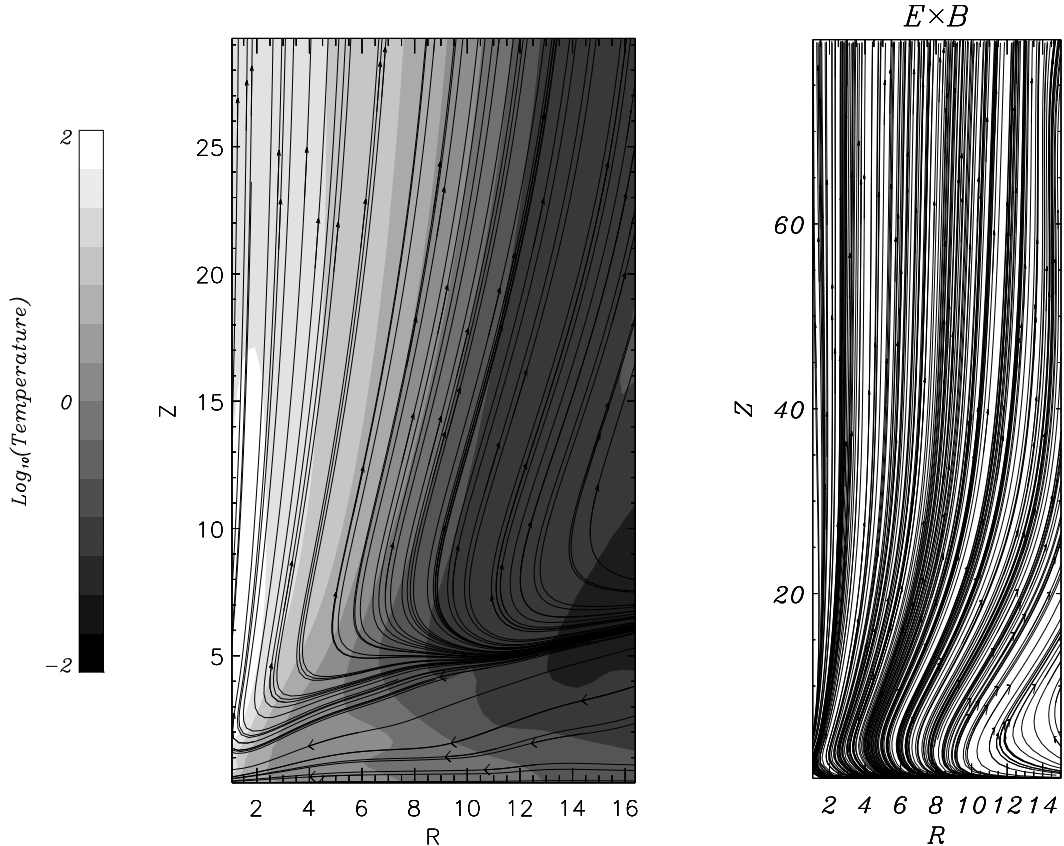}
\caption{Left: \pol flow streamlines and temperature. Both accretion and
  ejection regimes occur. The heated plasma (Ohmic heating and flow
  dynamics) in the disk give birth to a ``hot'' jet where temperature
  isocontours coincide with flow streamlines. The disk aspect ratio is
  small thanks to a predominant energy 
  transport achieved by jet launching rather than by enthalpy
  creation. Right: \pol 
  streamlines of the MHD Poynting flux. The main part of the MHD energy flux is
  sent into the jet and sustains powerful hot jets.}
\label{F6}
\end{figure*}

Nevertheless, these two last contributions are small compared to the
mechanical energy released by accretion so that $P_{LIB} \simeq 0.63
P_{MEC}$. This configuration is completely different from a usual
ADAF, since there one would expect to have $P_{LIB} \ll P_{MEC}$. 
The jet flow is completely
responsible for this seeming discrepancy. Indeed, 
as shown in the bottom-left plot
of Fig.~\ref{F5}, the power going into the jet (evaluated at $Z=25$)
is of the same order than
$P_{LIB}$ which leads to an energy budget which is indeed close to a
stationary state since  
$P_{LIB} = P_{MEC} + P_{ENT} + P_{MHD} \simeq P_{JET}$. The jet power
components $P_{MEC,J}$, $P_{ENT,J}$ and $P_{MHD,J}$ 
are shown in the bottom-right panel: the
mechanical energy flux is again the largest one but other components are
not negligible, in particular the jet MHD Poynting flux. This means that
in the corona, the magnetic field is still carrying energy and consequently
that acceleration of jet matter will still occur beyond this surface located at
$Z=25$. Note also that the enthalpy flux at the base of the jet is a
significant fraction of the jet power which confirms our previous statements
about the ``hot'' nature of the jet. As a final illustration of the energy
transfer that is dominated by the jet, we show in Fig.~\ref{F6} both \pol
streamlines and \pol MHD Poynting flux streamlines. We clearly see that
both accretion and ejection occurs and that the main part of the
Poynting flux is feeding the jet structure. Moreover, the disk
remains thin (as seen in the flow streamlines), although Joule heating
occurs within the disk. Here again, the role of the jet is primordial since
it enables the accretion disk to transfer the accretion energy into jet
power rather than in disk enthalpy (temperature) that would thicken
the disk. 
\section{Summary and outlook}

In this paper, we presented time-dependent MHD simulations of an
axisymmetric, resistive accretion disk threaded by a bipolar magnetic
field. The 
computation evolves the full set of MHD equations, including an 
energy equation with spatio-temporally varying resistivity triggered 
inside the disk only (set with an $\alpha$-type prescription). We assumed
that both the viscous torque and radiative 
losses were negligible, so that we can deal with radiatively inefficient,
magnetized accretion disks. Initial conditions were carefully
designed to fulfill the requirement of jet production, i.e. starting from
a near-equipartition accretion disk. Our computations bring important new results:
\begin{itemize}
\item{The present simulations are the first simulations to achieve a
near stationary state consisting of a magnetized accretion disk launching 
trans-Alfv\'enic up to super-fastmagnetosonic, collimated, ``hot''
  jets where thermal pressure balances magnetic effects. The initial
  configuration is general and valid for all sub-Eddington non-relativistic
  disks (modifications might occur in the very close vicinity of a black hole
  $R<10R_S$)  since it is
  assuming a Keplerian accretion disk 
  in a vertical hydrostatic equilibrium and threaded by a bipolar
  magnetic field. This result
  differs from our previous study (CK02) by the inclusion of a realistic energy
  equation replacing the simplifying polytropic relation. The obtained jet
  displays a temperature higher than the disk one, and is consistent
with the creation of a hot corona. In this jet, we have shown that a
significant part ($\sim 10\%$) of the jet power is enthalpy driven, 
 and the thermal pressure is of the order of the magnetic pressure. Moreover, 
the present
  simulations started from an initial magnetic configuration with
  bent magnetic field lines and therefore prove unambiguously 
that the obtained collimation of the
  jet is self-consistently built up. We have, in addition, completely
quantified the action of the magneto-centrifugal acceleration of matter,
by demonstrating the force balance along field lines as well as radially,
showing the butterfly current circuit and fully diagnosing the energy
  transfer.}
\item{Evaluating energy fluxes through inner and outer radius as well as
  through the disk surface, we presented the entire energy budget of the
  accretion-ejection structure. Unlike ADAF-like flows, we show that the
  main energy transport is achieved by the jet which channels the major part
  of the energy released by accretion. This kind of flow can lead to
  under-luminous accretion disks supporting bright jets (where the
  assumption of non-radiative plasma can no longer be valid), as for
  instance M87 \citep{Dima03}. In our simulations, the luminosity
  of the jet can be up to the energy liberated by the accretion $P_{JET}\sim
  GM_*\dot{M}_A/2R_I$, where $M_*$ is the mass of the central object and
  $R_I$ the inner radius of the jet. Therefore, these
  simulations are a complete model of the generic 
 ``Advection Dominated Inflow-Outflow Solutions'' (ADIOS) 
postulated by \citet{Blan99}, where the authors suggested that
  outflows might carry a part of the total disk energy. The accretion
  disk with its associated hot jet is no longer subject 
to strong enthalpy creation as in ADAF flow. This also implies that this
  kind of accretion disk can remain thin, even if some local heating 
(like the Ohmic heating included in the simulation) is
  occuring inside. Finally, these simulations adequately apply to both
  YSO and AGN type systems.}      
\end{itemize}
Forthcoming work will need to deal with alleviating the few simplifying
assumptions in this work, the most prominent being the
assumption of axisymmetry. This precludes our simulations from 
effects due to non-axisymmetric instabilities that might perturb the flow
\citep{Kim00} or the equipartition disk~\citep{Kep02}. We will be able
to check the stability of our simulations by stepping 
into a three-dimensional framework. Another
simplifying assumption was to suppose that the viscous 
torque is very small compared to the magnetic torque (as well as viscous
heating small compared to Joule heating). The inclusion of viscosity in our
simulations can lead to different kinds of flow where enthalpy creation can
be enhanced and the jet luminosity correspondingly decreased. 
The transfer of angular
momemtum will then be more complicated (both radial by viscosity and vertical by
the jet), but will undoubtedly enrich momentum, angular momentum,
and energy transport phenomena within the
accretion disk-jet system.    

\acknowledgments

The authors would like to thank the anonymous referee for helpful remarks
that have led to improve the clarity of the paper. 
This work was done under Euratom-FOM Association Agreement with
financial support from NWO, Euratom, and the European Community's Human
Potential Programme under contract HPRN-CT-2000-00153, PLATON, also
acknowledged by F.C. NCF is acknowledged for providing computing facilities.


\begin{thebibliography}{}
\bibitem[Blandford \& Begelman(1999)]{Blan99} Blandford, R.D. \& Begelman,
  M.C. 1999, \mnras, 303, L1
\bibitem[Blandford \& Payne(1982)]{Blan82} Blandford, R.D. \& Payne,
D.G. 1982, \mnras, 199, 883
\bibitem[Brackbill \& Barnes(1980)]{brack80} Brackbill, J.U. \& Barnes,
  D.C. 1980, J. Comput. Phys., 35, 426
\bibitem[Casse \& Ferreira(2000)]{Cass00a} Casse F. \& Ferreira J., 2000
\aap, 353, 1115
\bibitem[Casse \& Keppens(2002)]{cass02} Casse, F. \& Keppens, R. 2002,
\apj, 581, 988
\bibitem[DiMatteo et al.(2003)]{Dima03} Di Matteo ,T. et al. 2003, \apj,
  582, 133
\bibitem[DiMatteo et al.(2000)]{Dima00} Di Matteo, T. et al. 2000, \mnras,
  311, 507
\bibitem[Ferreira(1997)]{Ferr97} Ferreira J. 1997 \aap, 319, 340
\bibitem[Ferreira \& Pelletier(1995)]{Ferr95} Ferriera, J. \& Pelletier,
  G. 1995,\aap, 295, 807
\bibitem[Frank et al.(1992)]{Fran92} Frank, J., King, A. \& Raine,
  D. 1992, Accretion power in astrophysics (Cambridge University press) 
\bibitem[Hartigan et al.(1995)]{Hart95} Hartigan, P., Edwards, S. \&
Ghandour, L. 1995, \apj, 452, 736
\bibitem[Heyvaerts \& Norman(1989)]{Heyv89} Heyvaerts, J. \& Norman
C. 1989, \apj, 347, 1055
\bibitem[Igumenshchev, Narayan \& Abramowicz(2003)]{Igum03} Igumenshchev,
  I.V., Narayan, R. \& Abramowicz, M.A. 2003, \apj, in press, astro-ph/0301402
\bibitem[Kato et al.(2002)]{Kato02} Kato, S., Kudoh, T., Shibata,
K. 2002, \apj, 565, 1035
\bibitem[Kim \& Ostriker(2000)]{Kim00} Kim, W.T. \& Ostriker, E.C. 2000,
  \apj, 540, 372
\bibitem[Keppens, Casse \& Goedbloed(2002)]{Kep02} Keppens, R., Casse, F,
\& Goedbloed, J.P. 2002, \apjl, 569, L121
\bibitem[Krasnopolski et al.(1999)]{Kras99} Krasnopolski, R., Li, Z.Y. \&
Blandford, R.D. 1999, \apj, 526, 631
\bibitem[Lery et al.(1999)]{Lery99} Lery, T., Henricksen, R.N. \& Fiege
  J.D. 1999, \aap, 350, 254
\bibitem[Li(1996)]{Li96} Li, Z.Y. 1996, \apj, 465, 855
\bibitem[Livio(1997)]{Livi97} Livio, M. 1997, in IAU
Colloquium 163. ASP Conference Series; Vol. 121; eds. D. T. Wickramasinghe;
G. V. Bicknell; and L. Ferrario (1997), p.845
\bibitem[Lynden-Bell \& Pringle(1974)]{Lynd74} Lynden-Bell, D. \& Pringle,
  J.E. 1974, \mnras, 168, 603
\bibitem[Matsumoto et al.(1996)]{Mats96} Matsumoto, R. et al. 1996, \apj, 461, 115
\bibitem[Mirabel et al.(1998)]{Mira98} Mirabel, I.F., Dhawan, V., Chaty,
S., et al. 1998, \aap, 330, L9
\bibitem[Mouschovias(1976)]{Mous76} Mouschovias, T.C. 1976, \apj, 206, 753
\bibitem[Narayan \& Yi(1995)]{Nara95} Narayan, R. \& Yi, I. 1995, \apj,
  452, 710
\bibitem[Novikov \& Thorne(1973)]{Novi73} Novikov , I.D. \& Thorne,
  K.S. 1973, in Blackholes, ed. C. Dewitt \& B. Dewitt (New York, Gordon \&
  Breach), 343 
\bibitem[Ouyed \& Pudritz(1997)]{Ouye97} Ouyed, R. \& Pudritz, R. 1997,
\apj, 482, 712
\bibitem[Paczynsky \& Wiita(1980)]{Pacz80} Paczynsky, B. \& Wiita,
  P.J. 1980, \aap, 88, 23
\bibitem[Rees et al.(1982)]{Rees82} Rees, M.J., Begelman, M.C., Blandford,
  R.D. \& Phinney, E.S. 1982, \nat, 295, 17
\bibitem[Rekowski et al.(2003)]{Reko03} Rekowski, B.V., Branderburg, A.,
  Dobler, W. \& Shukurov, A. 2003, \aap, 398, 825
\bibitem[Rutten et al.(1992)]{Rutt92} Rutten, R.G.M., Van Paradijs, J. \&
  Tinbergen, J. 1992, \aap, 260, 213 
\bibitem[Sauty et al.(2002)]{Saut02} Sauty, C., Trussoni, E. \& Tsinganos,
K. 2002, \aap, 389, 1068
\bibitem[Serjeant et al.(1998)]{Serj98} Serjeant, S., Rawlings, S., Lacy,
M. et al. 1998, \mnras, 294, 494
\bibitem[Shakura \& Sunyaev(1973)]{Shak73} Shakura, N.I. \& Sunyaev,
  R.A. 1973, \aap, 24, 337
\bibitem[Shapiro et al.(1976)]{Shap76} Shapiro, S.L., Lightman, A.P. \&
  Eardley, D.M. 1976, \apj, 204, 187
\bibitem[T\'oth(1996)]{Toth96a} T\'oth, G. 1996, Astrophys. Lett., 34, 245
\bibitem[T\'oth \& Odstr\v cil(1996)]{Toth96b} T\'oth, G. \& Odstr\v cil,
  D. 1996, J. Comput. Phys., 128, 82
\bibitem[Ushida \& Shibata(1985)]{Ushi85} Ushida, Y. \& Shibata, K. 1985,
\pasj 37, 515
\bibitem[Ustyugova et al.(1995)]{Usty95} Ustyugova, G.V., Koldoba, A.V.,
Romanova, M.N., Chetchetkin, V.M. \& Lovelace, R.V.E. 1995, \apj, 439, L39
\bibitem[Ustyugova et al.(1999)]{Usty99} Ustyugova, G.V., Koldoba, A.V.,
Romanova, M.N., Chetchetkin, V.M. \& Lovelace, R.V.E 1999, \apj, 516, 221
\bibitem[Wardle \& K\"onigl(1993)]{Ward93} Wardle, M. \& K\"onigl, A. 1993,
\apj, 410, 218
\end{thebibliography}
\end{document}